\documentclass{appolb}
\usepackage{graphicx}
\usepackage{comment}
\usepackage{epsfig,latexsym,amssymb}
\usepackage{amsmath}
\usepackage{bm}
\usepackage{slashed}
\usepackage{subfigure}
\usepackage{color}

\newcommand{\be}{\begin{equation}}
\newcommand{\ee}{\end{equation}}
\newcommand{\bea}{\begin{eqnarray}}
\newcommand{\eea}{\end{eqnarray}}
\newcommand{\nn}{\nonumber}

\newcommand{\as}{\alpha_s}

\def\eq#1{{Eq.~(\ref{#1})}}

\newcommand{\ben}{\begin{eqnarray*}}
\newcommand{\een}{\end{eqnarray*}}

\begin{document}

\title{Centrality and energy dependence of charged 
particles in p+A and A+A collisions from running 
coupling $k_{T}$-factorization\thanks{Presented by
A.V.G. at Diffraction and Low-x 2018.} }
\author{Adrian Dumitru
\address{Department of Natural Sciences, Baruch College, CUNY,
17 Lexington Avenue, New York, NY 10010, USA \\
The Graduate School and University Center,
The City University of New York, 365 Fifth Avenue, New York, NY 10016, USA}\\
%
{Andre V. Giannini, Matthew Luzum}
\address{Instituto de F\'{i}sica, Universidade de S\~ao Paulo,
Rua do Mat\~ao 1371, 05508-090 S\~ao Paulo-SP, Brazil}\\
{Yasushi Nara}
\address{Akita International University, Yuwa, Akita-city 010-1292, Japan}
}

\maketitle
\begin{abstract}
We extend the numerical analysis of the energy and 
centrality dependence of particle multiplicities 
at midrapidity in high-energy p+A and A+A collisions from a 
running coupling $k_T$-factorization formula made in~\cite{Dumitru:2018gjm} 
by considering two unintegrated gluon distributions that were left out. 
While a good agreement with the experimental data in A+A collisions 
is achieved, improving the description of those observables 
in p+A collisions calls for a better understanding of the 
proton unintegrated gluon distribution at larger values of 
$x$ and also the use of a realistic impact parameter dependence. 
\end{abstract}

\section{Introduction}

Over the past years the Color Glass Condensate framework for particle production
has been applied with success to understand the DIS data at
HERA energies and hadron production in a broad region
of the phase space (from central to very forward rapidities) at RHIC
and LHC energies~\cite{Albacete:2014fwa}.
In particular, calculations of hadron production at
midrapidity
are based on the $k_{T}$-factorization approach, where
the expression for inclusive (small-$x$) gluon production
has originally been derived assuming a fixed coupling.
Despite that, some 
studies~\cite{Kharzeev:2001gp,Levin:2010dw,Tribedy:2010ab} have
included running coupling effects in their calculation by
just replacing $\as\to \as (Q^2)$ in the relevant expressions
and, then, fixing the momentum scale $Q^2$ later by hand.
Since this is an arbitrary procedure, it comes as no surprise
that distinct predictions found in the literature were
obtained assuming different prescriptions for fixing $Q^2$.

Although the results obtained following this procedure
do not depend strongly on how $Q^2$ is fixed, here we follow
ref.~\cite{Dumitru:2018gjm} and
employ the $k_T$-factorization formula for single-inclusive
(small-$x$) gluon production in the scattering of two valence quarks
derived in~\cite{Horowitz:2010yg}, which results from a resummation
of the relevant one-loop corrections into the running of the
coupling\footnote{Our notation follows ref.~\cite{Horowitz:2010yg}: ${\bm k}$
denotes the transverse momentum of the produced gluon while ${\bm q}$
and ${\bm k}-{\bm q}$ are the ``intrinsic'' transverse momenta from
the gluon distributions.},
\begin{eqnarray}\label{eq:rcktfact}\nn
  \frac{d^3 \sigma}{d^2 k \, dy} \, = \, N\,  \frac{2 \, C_F}{\pi^2} \,
  \frac{1}{{\bm k}^2} \, \int d^2q \, d^2b \, d^2b'\,
       {\overline \phi}_{h_1} ({\bm q}, y,{\bm b})
  \, {\overline \phi}_{h_2} ({\bm k} - {\bm q}, Y-y,{\bm b}-{\bm b'}) \times
  \, \\
  \frac{\as \left(
      \Lambda_\text{coll}^2 \, e^{-5/3} \right)}{\as \left( Q^2 \,
      e^{-5/3} \right) \, \as \left( Q^{* \, 2}\, e^{-5/3} \right)}\,. \,\,\,\,\,
\end{eqnarray}
\eq{eq:rcktfact} should be convoluted with a
with fragmentation function in order to yield
results at a hadronic level (this procedure also fix
the collinear infrared cutoff $\Lambda_\text{coll}^2$,
which should match the momentum scale of the
fragmentation function~\cite{KovchegovWeigert}).
However, as $p_{T}$-integrated multiplicities
are dominated by the soft region ($p_{T} \ll 1$ GeV)
we keep the simple model for the fragmentation function
used in~\cite{Dumitru:2018gjm}:
$D(z,\mu_{FF}^{2})\sim \delta(1-z)$. A change in the
fragmentation function would mainly change the normalization
of our results and can be absorbed into the
normalization factor $N$ (which also accounts for
''K-factors'' due to high-order corrections and
will be determined by comparison with experimental data).

The unintegrated gluon distribution (UGD) is given by 
\begin{equation}\label{eq:rc_ktglueA}
  {\overline \phi} ({\bm k}, y,{\bm b})  =  \frac{C_F}{(2 \pi)^3} \,
  \int d^2 r \, e^{- i {\bm k} \cdot {\bm r}} \ \nabla^2_r \,
  \mathcal{N}_{A} ({\bm r}, y,{\bm b})\,,
\end{equation}
and do not involve a factor of $1/\alpha_s(k^2)$ as
in the fixed coupling $k_{T}$-factorization formula;
these factors appear explicitly in \eq{eq:rcktfact} with the appropriate
scale\footnote{We refer to~\cite{Horowitz:2010yg} for
the full expression for the scale figuring in $\as(Q^2 e^{-5/3})$.}.
$\mathcal{N}_{A}({\bm r}, y,{\bm b})$ denotes the forward (adjoint)
dipole scattering amplitude at impact parameter ${\bm b}$. As previous
works~\cite{Albacete:2012xq}, a uniform gluon density within a proton was
assumed.

\begin{figure}[htb]
\begin{center}
\includegraphics[width=7.0cm]{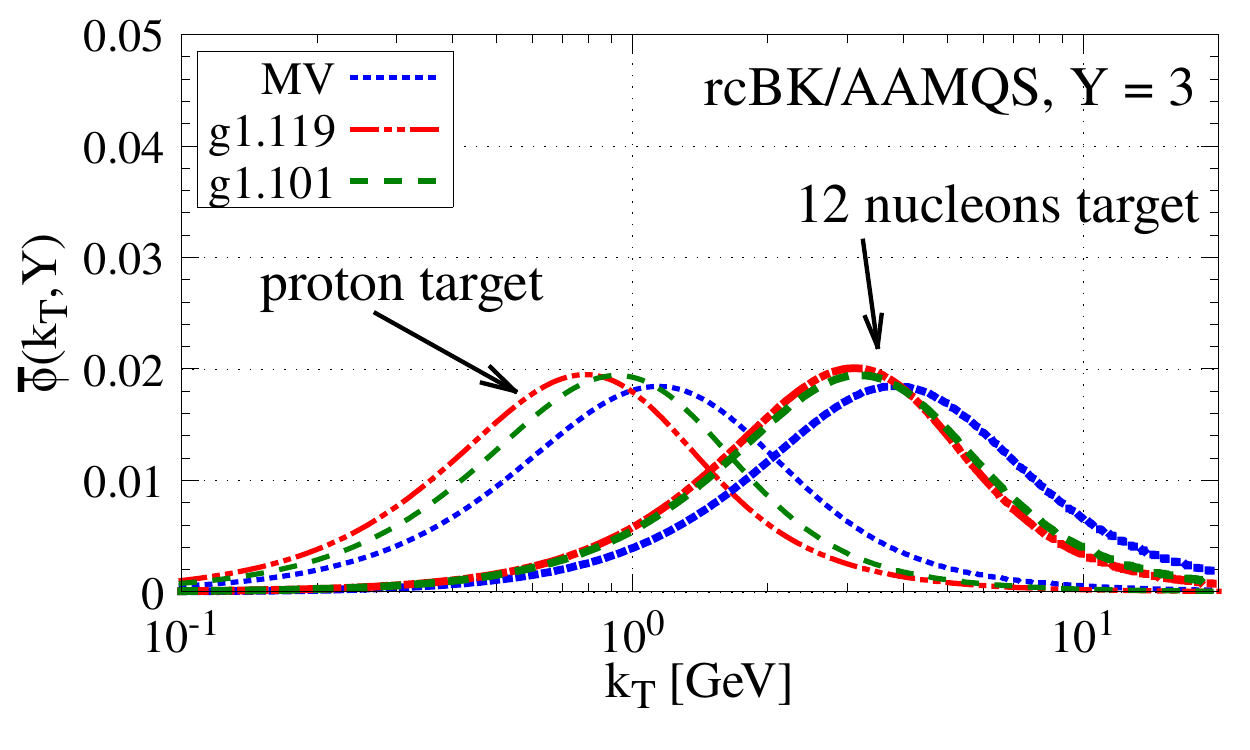}
\end{center}
\vspace*{-7mm}
\caption[a]{UGDs from different initial conditions
for rcBK equation at evolution rapidity $Y = 3$. The peak of
this function defines the saturation scale, $Q_{s}(Y)$.
}
\label{fig:phibarAllUGDs}
\end{figure}

As in~\cite{Dumitru:2018gjm,Albacete:2012xq}, $\mathcal{N}_{A}$
will be given by solutions of the running coupling
Balitsky-Kovchegov (rcBK) equation provided by the
AAMQS fits of the HERA data~\cite{AAMQS}.
However, while ref.~\cite{Dumitru:2018gjm} considered
only the McLerran-Venugopalan (MV) UGD set,
here we also consider the ``g1.119" and ``g1.101" UGD sets 
which are supposed to provide a better representation of the proton
UGD\footnote{We note that all these UGDs have already been
used to compute observables in hadronic collisions and
detailed information about them can be found in the discussion
around eq. (3) of the second work listed in ref.~\cite{Albacete:2012xq}.}.
Fig.~\ref{fig:phibarAllUGDs} shows the different
UGDs considered here for a proton and for a target made of 12 nucleons after three units of rcBK
rapidity evolution.


In what follows, we extend the analysis made in~\cite{Dumitru:2018gjm}
by presenting results for the energy and centrality dependence of
charged particle multiplicities produced in p+A and A+A collisions
from different rcBK evolved UGDs. The references for all experimental
data presented here can be found in~\cite{Dumitru:2018gjm}.


\section{Results, discussion and conclusions}

\begin{figure}[htb]
\begin{center}
\includegraphics[width=10.0cm]{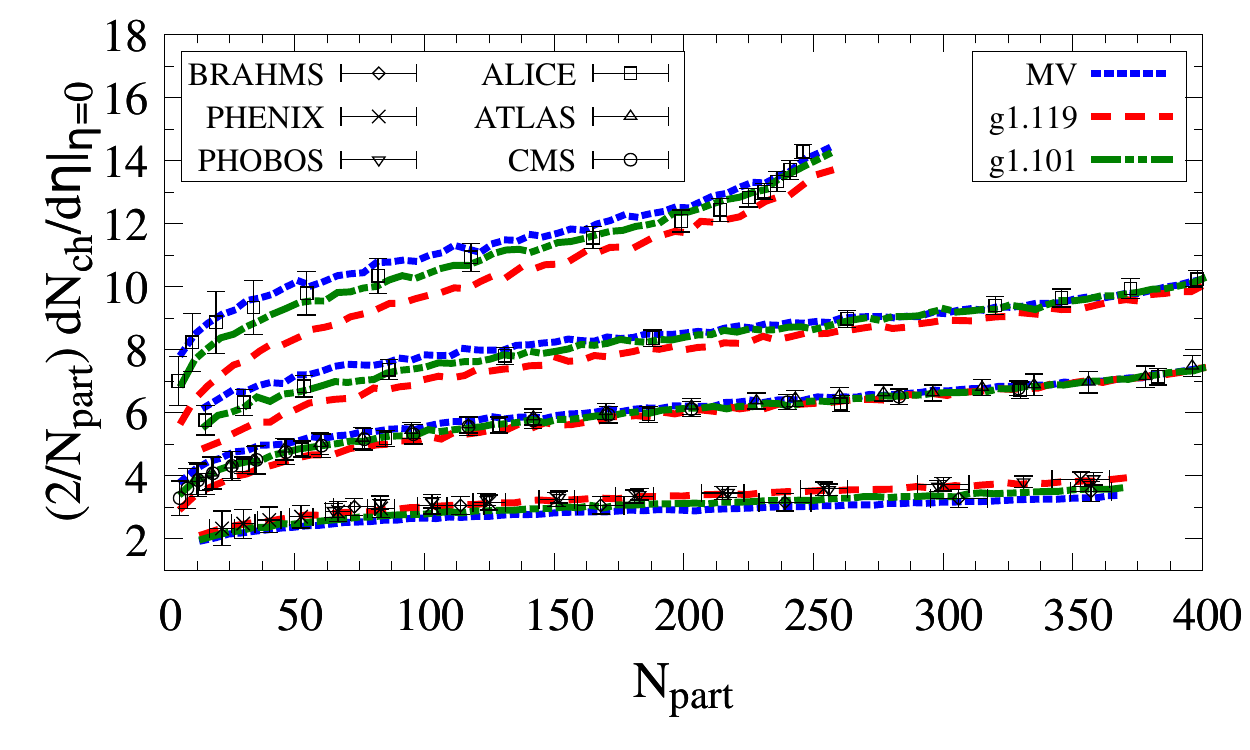}
\end{center}
\vspace*{-8.5mm}
\caption[a]{Left: Centrality dependence of the multiplicity per
  participant pair in Au+Au/Pb+Pb/Xe+Xe collisions at
  $\sqrt{s}=200$~GeV, 2.76~TeV, 5.02~TeV and 5.44~TeV.
  From bottom to top, curves and data points have been scaled by
  1.0/0.85/1.0/1.35 to improve visibility.}
\label{fig:AA_Npart}
\end{figure}

Following previous phenomenological 
works~\cite{Levin:2010dw,Tribedy:2010ab,Albacete:2012xq,Duraes:2016yyg}
we apply the $k_{T}$-factorization approach to compute 
the centrality and energy dependence of $dN_{ch}/d\eta$ in A+A 
collisions.
Figure~\ref{fig:AA_Npart} shows the results for the centrality 
dependence of the charged particle multiplicity in Au+Au 
and Pb+Pb/Xe+Xe collisions at RHIC and at LHC energies, respectively. 
The normalization figuring in~\eq{eq:rcktfact} has been fixed 
(for each UGD) by matching the central Pb+Pb data at 2.76 TeV; 
this same normalization has been used across all collision systems, 
energies and centralities considered.
One can see that while all UGDs present the well known increase 
of $dN_{ch}/d\eta$ per participant towards more central collisions
(which is related to the fact that the convolution of
the UGDs in~\eq{eq:rcktfact} increases as both transverse momentum
arguments can be near the ``saturation peak", thanks to A+A
collisions becoming more symmetric) the results with the g1.119
UGD become worse as the collision energy increases; on the other
hand, results with the g1.101 UGD compare well with the MV results
from~\cite{Dumitru:2018gjm} and describe the data within the
error bars.
%

\begin{figure}[htb]
\begin{center}
\includegraphics[width=7.8cm]{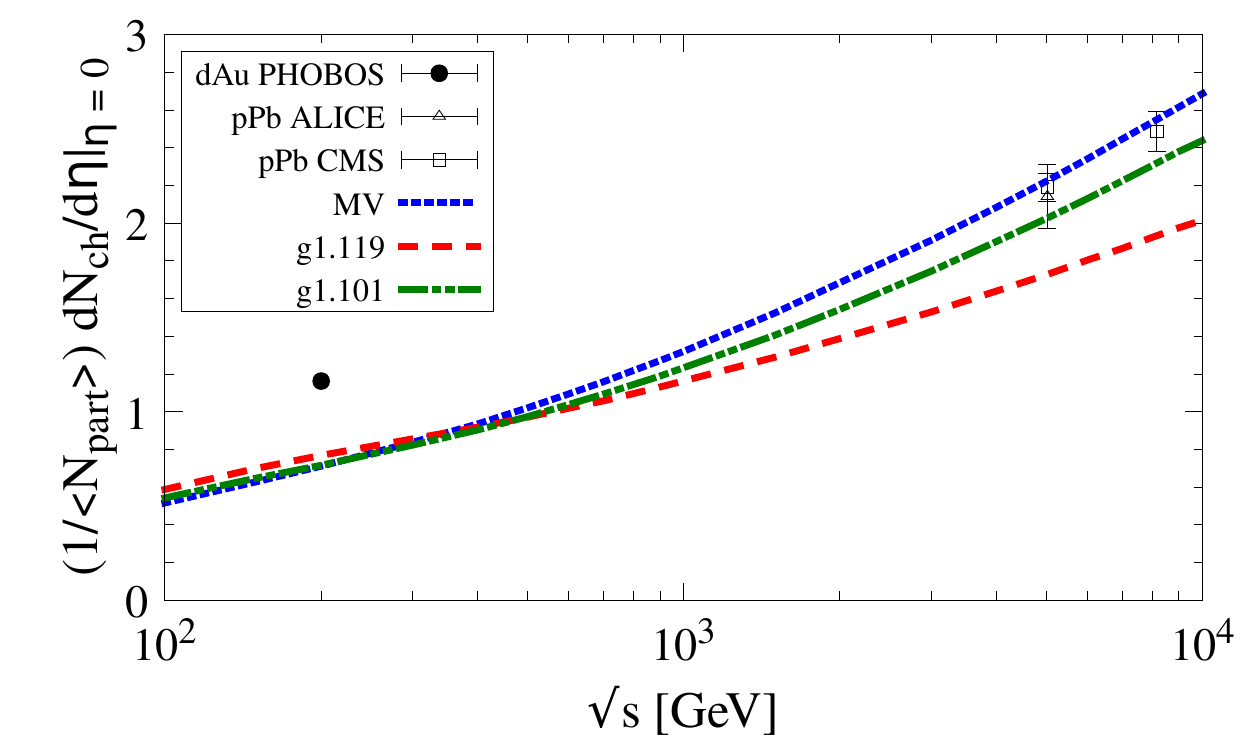}
\includegraphics[width=7.8cm]{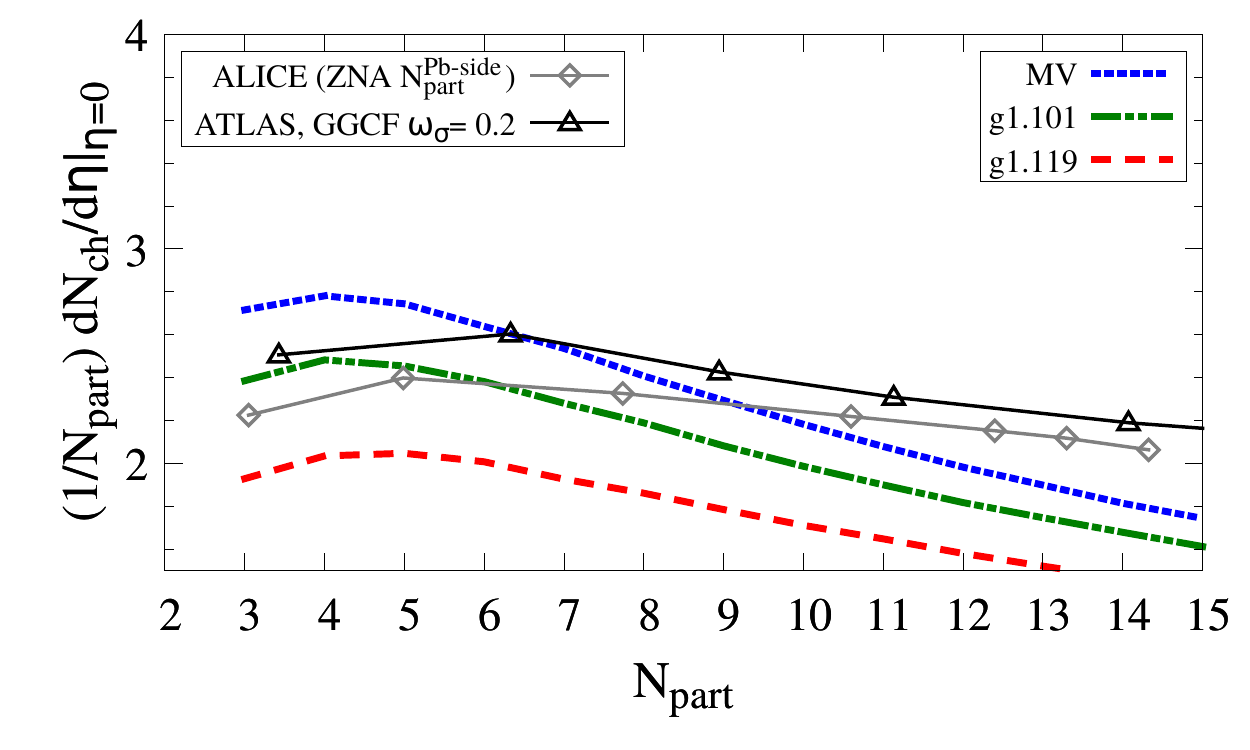}
\end{center}
\vspace*{-7mm}
\caption[a]{Energy and $N_{\rm{part}}$ dependence of the
charged particle multiplicity in p+Pb collisions at $\eta=0$.}
\label{fig:pA_roots_Npart}
\end{figure}

While we checked that all UGDs provide a similar description of
the energy evolution of the multiplicity per pair participant
in central (0-6\%) collisions\footnote{This fact can also be
inferred from Fig.~\ref{fig:AA_Npart} once all results presented
are at least in near accordance with the experimental
data for central collisions in all energy range considered.},
the situation in p+A collisions is more interesting. Fig.
\ref{fig:pA_roots_Npart} shows our results for the energy
and centrality dependence of the charged particle multiplicity
in p+Pb collisions.
While the results from the running coupling 
$k_{T}$-factorization formula for the energy dependence 
follow the same trend presented in Fig. \ref{fig:AA_Npart}
at LHC energies, all UGDs fail to describe the data at RHIC
top energy. This should be expected since in this case one is
sensitive mainly to the rcBK initial conditions (given at $x_{0}=0.01$),
rather than the small-$x$ evolution. The inclusion of additional corrections 
to \eq{eq:rcktfact}, as well as extending the CGC
framework to higher values of $x$~\cite{Jalilian-Marian:2017ttv}, 
could help to achieve a better understanding
of the proton UGD and lead to a better agreement 
with the data at RHIC energies.

The centrality dependence in p+Pb collisions is also interesting. 
We find that at $5.02$ TeV and beyond $N_{part} \simeq 4$ the 
multiplicity per participant actually decreases slightly 
with $N_{part}$, regardless the UGD considered. 
This is due to the fact that for increasingly {\it asymmetric} 
collisions (larger $N_{\rm part}$ for p+A) the 
convolution of the UGDs (in transverse momentum space)
does not increase in proportion to $N_{\rm part}$; 
a fit of the MV result for $5\le N_\text{part}\le
15$ gives $\sim \ln^{1.25}(N_{\rm part}) / N_{\rm part}$.
This same feature is also seen in the experimental data but 
with a somewhat flatter dependence on $N_{part}$. 
The origin of this difference can be related to the lack of 
a realistic impact parameter dependence of 
the proton-UGD in our computations and also 
due to the bias introduced in the experimental 
centrality selection. Results shown here can 
be improved by taking into account the bias on the 
configurations of the small-$x$ gluon fields 
via the reweighting procedure developed in~\cite{Dumitru:2018iko}, 
since the UGDs employed here have been averaged 
over all BK gluon emissions without any bias.

Here we extended the analysis of the energy and centrality 
dependence of the charged particle multiplicity produced 
at mid rapidities presented in~\cite{Dumitru:2018gjm} by 
considering two UGD sets that were left out in that first analysis. While 
two of them (MV and g1.101) provide a good description of the centrality and 
energy dependence in A+A collisions, for p+A collisions 
all UGDs are only in qualitative agreement with the 
$N_{\rm part}$ dependence measured at LHC energies
and fail to describe the RHIC data for the energy
evolution. Improving the agreement between theory and data 
in small collision systems and lower energies calls for 
a better understanding of the proton-UGD at higher values of 
$x$ and the inclusion of a realistic impact 
parameter dependence.

\section*{Acknowledgments}
A.D.\ acknowledges support by the DOE Office of Nuclear
Physics through Grant No.\ DE-FG02-09ER41620; and from The City
University of New York through the PSC-CUNY Research grant 60262-0048.
A.V.G. acknowledges support by the Brazilian funding agency FAPESP
through grant 17/14974-8.  M.L.~acknowledges
support from FAPESP projects 2016/24029-6 and 2017/05685-2, and
project INCT-FNA Proc.~No.~464898/2014-5.

\end{document}